\UseRawInputEncoding
\pdfoutput=1

\documentclass[showpacs,nofootinbib,showkeys,eqsecnum,prd,aps,preprint]{revtex4-2}

\usepackage[english]{babel}
\usepackage{comment}

\usepackage{amssymb}
\usepackage{amsmath}
\usepackage{amsfonts}
\usepackage{latexsym}
\usepackage{mathrsfs}
\usepackage{bm}
\usepackage{tensor}
\usepackage{hyperref}
\usepackage{graphicx}

\def\const{\mathop{\rm const}\nolimits\,}
\def\dac{\displaystyle\frac}
\def\dil{\displaystyle\int\limits}

\usepackage{amsthm}
\usepackage{xcolor}
\newtheorem{theorem}{Theorem}[section]

\newtheorem{corollary}[theorem]{Corollary}
\newtheorem{defin}{Definition}[section]
\newtheorem{prop}{Proposition}[section]

\usepackage{diagbox}

\def\llan{\langle\!\langle}
\def\rran{\rangle\!\rangle}

\def\pa{\partial}

\def\BFC{{\bf C}}

\def\const{\mathop{\rm const}\nolimits\,}

\def\dac{\displaystyle\frac}

\def\dil{\displaystyle\int\limits}
\def\{{\lbrace}
\def\}{\rbrace}

\def\Or{{\rm O}}

\begin{document}

\title{Semiclassical states localized on a one-dimensional manifold and governed by the nonlocal NLSE with an anti-Hermitian term}

\author{Anton E. Kulagin}
\email{aek8@tpu.ru}
\affiliation{Tomsk Polytechnic University, 30 Lenina av., 634050 Tomsk, Russia}
\affiliation{V.E. Zuev Institute of Atmospheric Optics, SB RAS, 1 Academician Zuev Sq., 634055 Tomsk, Russia}

\author{Alexander V. Shapovalov}
\email{shpv@mail.tsu.ru}
\affiliation{Department of Theoretical Physics, Tomsk State University, Novosobornaya Sq. 1, 634050 Tomsk, Russia}

\begin{abstract}
We develop the method for constructing solutions to the nonlocal nonlinear Schr\"{o}dinger equation (NLSE) with an anti-Hermitian term that are semiclassically localized on a one-dimensional manifold (a curve). The evolution of the curve is given by the closed system of integro-differential equations that can be treated as the "classical"\, analog of the open quantum system with the nontrivial geometry. Using our approach, we consider the evolution of vortex states in the open quantum system described by the specific model NLSE. The semiclassical stage of the vortex evolution can be treated as a quasi-steady vortex state. We show that the behaviour of this state is largely determined by the geometry of the localization curve.\\
\end{abstract}


\keywords{semiclassical approximation; vortex state; nonlocal nonlinearity; Maslov complex germ method; non-Hermitian operator; open quantum system}

\maketitle

\section{Introduction}
\label{sec0}
The generalized nonlinear Schr\"{o}dinger equations (NLSEs) with the nonlocal cubic nonlinearity describe the broad classes of dynamic regimes and steady states of Bose--Einstein condensates (BECs) in the mean field representation \cite{pitaevskii1999,leggett2001}. The nonlocal term in the equation is responsible for the long-range interaction that is essential, for example, in the condensates with the dipole-dipole one \cite{baranov2008,malomed2009,klaus2022,zhao2021,ribeiro23}. The anti-Hermitian contribution to the equation operator allows one to model BEC as an open system involving environmental impact \cite{ashida20,tarasov08,fetter01,arecchi2000,zezyulin10}.

Studies of the BEC with the generalized NLSE, especially nonlocal one and involving non-Hermitian terms, are usually performed using numerical methods due to the mathematical complexity of the model \cite{gpelab15,alfimov23,tsubota02}. Yet, it is possible to develop analytical or semi-analytical methods even for rather complex models if the variety of BEC states under consideration is limited to more specific one. In particular, the semiclassical formalism based on Maslov's theory \cite{Maslov2,BeD2} allows one to construct solutions to the model NLSE that inherit some features of the "classical"\, behavior. Such feature for the Maslov complex germ method is that the semiclassical solution is localized in a neighbourhood of the point that moves in the phase space of the dynamical system which plays the role of "classical equations". In case of the common linear Schr\"{o}dinger equation, such system is literally the classical Hamilton equations \cite{bagrov1}. However, the attractive feature of this method is that it can be applied to the nonlocal nonlinear problem \cite{shapovalov:BTS1}. The role of "classical equation"\, is fulfilled by the dynamical system for the moments of the desired solution in this case and this system substantially determines the qualitative behaviour of such respective asymptotic solutions.

Let us emphasize again that Maslov's semiclassical formalism, based on the theory of pseudo-differential operators, is quite flexible tool that can be well expanded and generalized for various kinds of nonlocal nonlinear evolution models and solution classes. Like this, in \cite{kulq25}, the approach to constructing the solutions for the nonlocal NLSE that are localized in a neighbourhood of a finite number of trajectories was proposed. Such solutions are treated as the nonlinear superposition of "semiclassical quasiparticles"\, each of which is associated with its "classical trajectory"\, in the phase space. Since the localization points for every fixed time are distant from each other in the coordinate space, such approach is suitable for consideration of long-range interactions. However, such localization domain is still zero-dimensional and some matter states, e.g., the vortex one, could not be considered within that approach. Thus, it is of interest to generalize this formalism to the finite-dimensional localization manifolds.

In this work, we propose the method for constructing the solutions to the nonlocal NLSE with an anti-Hermitian terms that are semiclassically localized on a one-dimensional manifold (a curve) evolving over time. The evolution of the manifold is governed by the dynamical system that is determined by the NLSE. We will construct the solutions that belong to the class similar to the one used in \cite{LST16} for the asymptotic solutions to the nonlocal Fisher--Kolmogorov--Petrovskii--Piskunov equation and in \cite{sym2020} for the more specific nonlocal NLSE without anti-Hermitian terms (closed quantum system). The generalization of the last one to the open quantum systems provides the essential advancement in studying the nonstationary phenomena in BEC. In particular, it allows us to address the problem of description of the transient processes in the vortex states that lead to the formation of the vortex lattices \cite{tsubota02}. We consider this problem in the second part of the paper for one specific model within the proposed general formalism. The basic idea of our approach is a lift to the expanded space with a higher dimension where the asymptotics can be algorithmically constructed. Then the asymptotic solution of the NLSE can be obtained as the projection of the solution in the expanded space to the original one.

The paper is organized as follows. In Section \ref{sec1}, we pose the problem and give the formal definition of the semiclassical localization. In Section \ref{sec2}, we derive the formal asymptotic expansion of the desired solution to the Cauchy problem for the generalized NLSE using the lift to the space with a higher dimension. The semiclassical ansatz is introduced. In Section \ref{sec3}, we derive the explicit algebraic restriction to the class of functions where the asymptotic solutions exist. In Section \ref{sec:example}, the formalism proposed is applied to the specific model equation that describes the formation of vortex lattices. The concept of the quasi-steady vortex state is introduced that can appear during transient processes in the open quantum system. The geometrical interpretation of the dynamics for such state is given. In Section \ref{sec:con}, we conclude with some remarks.

\section{Nonlocal NLSE with an anti-Hermitian term. Classical equations}
\label{sec1}

The general form of the non-Hermitian nonlocal NLSE under consideration that reads as follows:
\begin{equation}
\begin{array}{l}
\bigg\{ -i\hbar\pa_t + H(\hat{z},t)[\Psi]-i\hbar \Lambda \breve{H}(\hat{z},t)[\Psi] \bigg\}\Psi(\vec{x},t)=0, \cr \cr
H(\hat{z},t)[\Psi]=V(\hat{z},t)+\varkappa\dil_{{\mathbb{R}}^n}d\vec{y}\,\Psi^{*}(\vec{y},t)W(\hat{z},\hat{w},t)\Psi(\vec{y},t), \cr
\breve{H}(\hat{z},t)[\Psi]=\breve{V}(\hat{z},t)+\varkappa\dil_{{\mathbb{R}}^n}d\vec{y}\,\Psi^{*}(\vec{y},t)\breve{W}(\hat{z},\hat{w},t)\Psi(\vec{y},t).
\end{array}
\label{hartree1}
\end{equation}
Here, $\vec{x}\in {\mathbb{R}}^n$, $\hat{\vec{p}}_x=-i\hbar\pa_{\vec{x}}$, $\hat{z}=(\hat{\vec{p}}_x,\vec{x})$, $\hat{w}=(\hat{\vec{p}}_y,\vec{y})$, and $\hbar$ is a formal small asymptotic parameter. As usual for the semiclassical formalism, an operator of the equation is defined in terms of pseudo-differential operators \cite{Maslov1,maslov81}. So, the operators $V(\hat{z},t)$, $\breve{V}(\hat{z},t)$, $W(\hat{z},\hat{w},t)$, $\breve{W}(\hat{z},\hat{w},t)$ belong to the set ${\mathcal{A}}^t_{\hbar}$ of pseudo-differential operators with smooth symbols growing not faster than polynomial (some related properties and formal definitions are given in Appendix \ref{app0}). Since the equation \eqref{hartree1} is given in the coordinate representation, the operator $\vec{x}$ is the operator of multiplication by $\vec{x}$. Hereinafter, we put a right arrow over and only over the $n$-dimensional vectors.

\begin{defin}
The function $\Psi(\vec{x},t,\hbar)$ belongs to the class ${\mathcal{T}}^t_\hbar\left(Z(s,t),\mu(s,t)\right)$, $s\in {\mathbb{D}}\subset {\mathbb{R}}$, of functions semiclassically concentrated on the curve $z=Z(s,t)$ with a weight $\mu(s,t)$ for every fixed $t\in[0,T]$ if for any operator $\hat{A}\in{\mathcal{A}}_{\hbar}^{t}$ with a Weyl symbol $A(z,t,\hbar)$ the followings holds:
\begin{equation}
\begin{gathered}
\lim\limits_{\hbar\to 0}\langle\Psi | \hat{A} | \Psi \rangle (t,\hbar)=\dil_{{\mathbb{D}}}\mu(s,t) A(Z(s,t),t,0)ds \quad \forall t\in[0,T], \\ \langle\Psi | \hat{A} | \Psi \rangle (t,\hbar)=\dil_{{\mathbb{R}}^n}\Psi(\vec{x},t,\hbar)\hat{A} \Psi^*(\vec{x},t,\hbar)d\vec{x}.
\end{gathered}
\label{def1a}
\end{equation}
\end{defin}
Hereinafter, $z\in {\mathbb{R}}^{2n}$. We will consider only the case when ${\mathbb{D}}=[s_1,s_2]$ and $s_1$, $s_2$ do not depend on $\hbar$. Let us also denote the $t$-parameterized curve under consideration as $\Lambda_t$ that formally reads
\begin{equation}
\Lambda_t=\left\{z=Z(s,t)\Big| s\in[s_1,s_2]\right\}.
\end{equation}

Let $\mu_{\Psi}(t,\hbar)=\langle\Psi|\Psi\rangle(t,\hbar)=||\Psi||^2(t,\hbar)$. From \eqref{def1a}, it is clear that the following relation holds in the class ${\mathcal{T}}^t_\hbar\left(Z(s,t),\mu^{(0)}(s,t)\right)$:
\begin{equation}
\lim\limits_{\hbar\to 0}\mu_{\Psi}(t,\hbar)=\dil_{{\mathbb{D}}}\mu^{(0)}(s,t)ds.
\label{limsig1}
\end{equation}
We will denote the second functional parameters of the class ${\mathcal{T}}^t_\hbar\left(Z(s,t),\mu^{(0)}(s,t)\right)$ by $\mu^{(0)}(s,t)$ since it corresponds to the zeroth order approximation (with respect of $\hbar$) of the zeroth moments in a sense that will be clarified later.


Obviously, the functional parameters $Z(s,t)$ and $\mu^{(0)}(s,t)$ of the class ${\mathcal{T}}^t_\hbar\left(Z(s,t),\mu^{(0)}(s,t)\right)$ can not be arbitrary if we require $\Psi(\vec{x},t)$ to be a solution to \eqref{hartree1}. For consistency of \eqref{def1a} with \eqref{hartree1}, we subject these functions to the following system:

\begin{equation}
\begin{gathered}
\dot{\mu}^{(0)}(s,t)=-2\Lambda\mu^{(0)}(s,t)\left( \breve{V}(Z(s,t),t)+\varkappa \dil_{{\mathbb{D}}}dr\,\mu^{(0)}(r,t)\breve{W}(Z(s,t),Z(r,t),t)\right), \\
\dot{Z}(s,t)=JV_z(Z(s,t),t)+\varkappa\dil_{{\mathbb{D}}}dr\, \mu^{(0)}(r,t)J W_z(Z(s,t),Z(r,t)).
\end{gathered}
\label{hes1}
\end{equation}
We term equations \eqref{hes1} as the zeroth order Hamilton-Ehrenfest system (HES) by analogy with \cite{shapovalov:BTS1}. The derivation of \eqref{hes1} is given in Appendix \ref{app0b}.

Note that not all well-posed initial conditions for \eqref{hes1} correspond to any continuous initial conditions for \eqref{hartree1}. For such correspondence, we should put the Bohr--Sommerfeld quantization condition on $Z(s,0)$ that reads
\begin{equation}
\begin{gathered}
\dil_{s_1}^{s_2} \langle \vec{P}(s,0),\vec{X}_s(s,0)\rangle ds= N\hbar, \quad N\in {\mathbb{Z}},
\end{gathered}
\label{p2hes11}
\end{equation}
where $\vec{X}_s(s,t)=\dac{\pa \vec{X}(s,t)}{\pa s}$. The notation $\langle \cdot,\cdot \rangle$ stands for the Euclidean scalar product of vectors. 

\begin{prop} \label{statement1} If $\Lambda_t$ is a closed curve, $N$ is conserved on solutions to \eqref{hes1}, i.e. we have
\begin{equation}
\begin{gathered}
\dil_{s_1}^{s_2} \langle \vec{P}(s,t),\vec{X}_s(s,t)\rangle ds= \const=N\hbar.
\end{gathered}
\label{p2hes11}
\end{equation}
\end{prop}
The proof is given in Appendix \ref{app0c}. Note that it is valid even for $\Lambda\neq 0$. We will give a physical meaning to it in Section \ref{sec:example}.

\section{Class of functions. Asymptotic expansion}
\label{sec2}
Let us introduce the auxiliary family of classes of trajectory concentrated functions ${\mathcal P}_\hbar^{s,t}(Z(s,t), S(s,t))$ that reads as follows:
\begin{align}
{\mathcal{P}}_\hbar^{s,t}(Z(s,t), S(s,t))=\bigg\{
\Phi:\Phi(\vec{x},s,t,\hbar)=\hbar^{-(n-1)/4}\cdot\varphi\Big(\frac{\Delta\vec{x}}{\sqrt{\hbar}},
s,t,\hbar\Big) \cdot\exp\Big[\frac{i}{\hbar}\left(S(s,t)+\langle \vec{P}(s,t),\Delta \vec{x}  \rangle\right) \Big] \bigg\}.
\label{pth1}
\end{align}
Here, $\Phi(\vec{x},s,t,\hbar)$ is a general element of the class ${\mathcal{P}}_\hbar^{s,t}(Z(s,t), S(s,t))$; the real functions $Z(s,t)=(\vec{P}(s,t),\vec{X}(s,t))$ and $S(s,t)$ are functional parameters of the class ${\mathcal{P}}_\hbar^{s,t}(Z(s,t), S(s,t))$; $\Delta\vec{x}=\vec{x}-\vec{X}(s,t)$; the functions $Z(s,t), S(s,t)$, and $\varphi(\vec{\xi},s,t,\hbar)$ smoothly depend on $t$ and $s$, regularly depend on $\sqrt{\hbar}$ in a neighbourhood of $\hbar=0$. The rigorous description of the function $\varphi(\vec{\xi},s,t,\hbar)$ will be given later in Section \ref{sec3}. Now it is enough to assume that $\varphi(\vec{\xi},s,t,\hbar)$ depends on $\vec{\xi}$ in such a way that the latter equations have a sense. The class \eqref{pth1} is the $s$-parameterized version of the class of trajectory concentrated functions that was used, e.g., in \cite{shapovalov:BTS1}. Here,  for brevity, we will denote it as ${\mathcal{P}}_\hbar^{s,t}$ where it does not cause the confusion.

Hereinforth, the functional parameters $Z(s,t)$ of the class ${\mathcal{P}}_\hbar^{s,t}$ will be subjected to the system of equation \eqref{hes1}, and the functional parameter $S(s,t)$ will be given by the following relation:
\begin{equation}
\begin{gathered}
\dot{S}(s,t)=\langle \vec{P}(s,t),\dot{\vec{X}}(s,t) \rangle - V(Z(s,t),t)-\varkappa \dil_{{\mathbb{D}}}dr\, \mu^{(0)}(r,t)W(Z(s,t),Z(r,t),t).
\end{gathered}
\label{moms10}
\end{equation}
One can readily verify that the function \eqref{moms10} satisfies the following equation on solutions to \eqref{hes1}:
\begin{equation}
S_s(s,t)=\langle \vec{P}(s,t),\vec{X}_s(s,t)\rangle.
\label{abbac1}
\end{equation}
Hence, the function \eqref{moms10} is somewhat analog of the abbreviated classical action.

We will seek for a solution to the equation \eqref{hartree1} in the class of functions that is defined in terms of \eqref{pth1} as follows:
\begin{equation}
\begin{gathered}
{\mathcal{J}}_\hbar^{\tau,t}(Z(\cdot,t), S(\cdot,t))=\bigg\{
\Psi:\Psi(\vec{x},t,\hbar)=\Phi(\vec{x},s,t,\hbar)\Big|_{s=\tau(\vec{x},t)} \bigg\}, \\
\Phi(\vec{x},s,t,\hbar)\in{\mathcal{P}}_\hbar^{s,t}(Z(s,t), S(s,t)).
\end{gathered}
\label{pth2}
\end{equation}
The class ${\mathcal{J}}_\hbar^{\tau,t}(Z(\cdot,t), S(\cdot,t))$ has one additional functional parameter $\tau(\vec{x},t)$. We will determine the family of hypersurfaces $s=\tau(\vec{x},t)$ by the following equation:
\begin{equation}
\langle \vec{X}_s(s,t),\Delta \vec{x}\rangle=0.
\label{hyper1}
\end{equation}
The solution to \eqref{hyper1} is chosen in such a way that the function $\tau(\vec{x},t)$ is continuous with respect to $\vec{x}$ and $t$.

Thus, \eqref{pth2} is the class of functions that are projections of $s$-parametrized trajectory concentrated functions \eqref{pth1}.

\begin{theorem} \label{pros-t1}
The following asymptotic estimates hold in the class \eqref{pth2} \cite{sym2020}:\vspace{-3pt}
\begin{eqnarray}
& \{\Delta\hat z\}^\alpha=\hat{\Or}(\hbar^{|\alpha|/2}), \quad
\Delta\hat z=(\Delta\hat{\vec p},\Delta\vec x).\label{pros6}
\end{eqnarray}
Here $\{\Delta{\hat z}\}^\alpha$ is the operator with the Weyl symbol $(\Delta z)^\alpha$,\vspace{-3pt}
\begin{eqnarray*}
& \Delta \hat{z}=(\Delta\hat{\vec{p}},\Delta\vec x),\\
& \Delta\hat{\vec{p}}=\hat{\vec{\pi}}-\vec P(\tau(\vec x,t),t), \quad \Delta\vec x=\vec
x-\vec X(\tau(\vec x,t),t ),\vspace{-3pt}
\end{eqnarray*}
$\alpha\in{\mathbb Z}^{2n}_+$ is a $2n$-dimensional multiindex ($2n$-tuple),  $\alpha=({\alpha_1}, \dots , {\alpha_{2n}})$, $\alpha_j = \overline{0,\infty}$, $j=\overline{1,2n}$:
\begin{equation}
v^{\alpha}=v_1^{\alpha_1}v_2^{\alpha_2}\cdot...\cdot v_{2n}^{\alpha_{2n}}, \quad
|\alpha|=\alpha_1+\alpha_2+...+\alpha_{2n}, \quad \frac{\partial^{|\alpha|}}{\partial z^{\alpha}}=
\prod_{j=1}^{2n}\frac{\partial^{\alpha_j}}{\partial z_j^{\alpha_j}}.
\label{multiind1}
\end{equation}
The operator $\hat{\vec{\pi}}$ acts as follows:
\begin{equation}
\hat{\vec{\pi}}\Psi(\vec{x},t)=-i\hbar\pa_{\vec{x}}\Phi(\vec{x},s,t)\Big|_{s=\tau(\vec{x},t)}.
\label{piop1}
\end{equation}
\end{theorem}
As mentioned earlier, all estimates throughout the text are uniform in time for $t\in[0,T]$.

\begin{corollary} The following asymptotic estimates hold in the class \eqref{pth1} \cite{sym2020}:\vspace{-3pt}
\begin{equation}
\Delta x_j=\hat{\Or}(\sqrt\hbar), \quad \Delta\hat p_j=\hat{\Or}(\sqrt\hbar), \quad
j=\overline{1,n},\vspace{-3pt}\label{pros16}
\end{equation}
and\vspace{-3pt}
\begin{equation}
\begin{gathered}
\{-i\hbar\partial_s- \langle{\vec
P}_s(s,t), \Delta\vec x\rangle+\langle{\vec X}_s(s,t), \Delta\hat{\vec
p}\rangle\}\Big|_{s=\tau(\vec x,t)}= \hat{\Or}(\hbar),\\
\Big\{-i\hbar\partial_t+V(Z(s,t),t)+ \varkappa \dil_{{\mathbb{D}}}dr\, \mu^{(0)}(r,t)W(Z(s,t),Z(r,t),t) +\\
+ \langle V_{z}(Z(s,t),t), \Delta\hat{z}\rangle+\varkappa \dil_{{\mathbb{D}}}dr\, \mu^{(0)}(r,t)\langle W_z(Z(s,t),Z(r,t),t),\Delta \hat{z}\rangle\Big\}\Big|_{s=\tau(\vec x,t)}= \hat{\Or}(\hbar).
\end{gathered}
\label{pros17}
\end{equation}
\end{corollary}

From \eqref{hyper1}, we also have
\begin{equation}
\begin{array}{c}
\tau_{\vec{x}}(\vec x,t)=\displaystyle\frac 1{{\vec X}{}_s^2(s,t)- \langle {\vec
X_{ss}}(s,t),\Delta\vec x\rangle}{\vec X}_s(s,t) \Big|_{s=\tau(\vec
x,t)},\cr \cr
\tau_t(\vec x,t)=\displaystyle\frac{- 1}{{\vec
X}{}_s^2(s,t)- \langle {\vec X_{ss}}(s,t),\Delta\vec x\rangle}\Big[\langle{\vec
X}{}_s(s,t),{\vec X_t}(s,t)\rangle- \langle{\vec X_{st}}(s,t),\Delta\vec
x\rangle\Big] \Big|_{s=\tau(\vec x,t)},
\end{array}
\label{asS5}
\end{equation}
that yields the following expansions:
\begin{equation}
\begin{array}{l}
\tau_{\vec{x}}(\vec x,t)=\bigg\{\displaystyle\frac{{\vec
X}_s(s,t)}{{\vec X}_s{}^2(s,t)} \bigg[1+\frac {\langle{\vec
X_{ss}}(s,t), \Delta\vec
x\rangle}{{\vec X}{}_s^2(s,t)}+\bigg(\displaystyle\frac {\langle{\vec X_{ss}}(s,t),\Delta\vec x\rangle}{{\vec
X}{}_s^2(s,t)} \bigg)^2\bigg]\bigg\}\bigg|_{s=\tau(\vec
x,t)}+ \cr
+\Or (\hbar^{3/2}), \cr \cr
\tau_t(\vec x,t)=\bigg\{-\displaystyle\frac{\langle{\vec X}{}_s(s,t),{\vec
X_t}(s,t)\rangle}
 {{\vec X}_s{}^2(s,t)} \bigg[1-\displaystyle\frac{\langle{\vec X_{st}}(s,t),\Delta\vec x\rangle}
 {\langle{\vec X}{}_s(s,t),{\vec X_t}(s,t)\rangle} +   \frac {\langle{\vec X_{ss}}(s,t),
\Delta\vec x\rangle}{{\vec X}{}_s^2(s,t)}+\cr
 +\bigg(\displaystyle\frac
{\langle{\vec X_{ss}}(s,t),\Delta\vec x\rangle}{{\vec X}{}_s^2(s,t)}
 \bigg)^2-\displaystyle\frac{\langle{\vec
X_{st}}(s,t),\Delta\vec x\rangle}{\langle{\vec X}{}_s(s,t),{\vec
X_t}(s,t)\rangle}\frac {\langle{\vec X_{ss}}(s,t), \Delta\vec x\rangle}{{\vec
X}{}_s^2(s,t)}\bigg]\bigg\}\bigg|_{s=\tau(\vec
x,t)}+\Or (\hbar^{3/2}).
\end{array}
\label{asS6}
\end{equation}

Let us denote the operator of equation \eqref{hartree1} as follows:
\begin{equation}
\hat{L}[\Psi]=-i\hbar\pa_t + H(\hat{z},t)[\Psi]-i\hbar \Lambda \breve{H}(\hat{z},t)[\Psi].
\label{nlseop1}
\end{equation}

We will consider the operator $\hat{L}$ in the expanded space that is complemented by the curve parameter $s$, i.e. we will consider equation \eqref{hartree1} as equation with variables $(s,t,\vec{x})$ instead of $(t,\vec{x})$. It can be naturally done in the class ${\mathcal{J}}_\hbar^{\tau,t}$ and will allow us to use properties of the class ${\mathcal{P}}_\hbar^{s,t}$. In order to write the operator $\hat{L}[\Phi]$ in the expanded space, we should prolong partial derivatives as follows:
\begin{equation}
\begin{gathered}
\pa_t \Psi(\vec{x},t)=\Big(\pa_t + \tau_t(\vec{x},t)\pa_s\Big)\Phi(\vec{x},s,t)\Big|_{s=\tau(\vec{x},t)},\\
\pa_{\vec{x}} \Psi(\vec{x},t)=\Big(\pa_{\vec{x}} + \tau_{\vec{x}}(\vec{x},t)\pa_s\Big)\Phi(\vec{x},s,t)\Big|_{s=\tau(\vec{x},t)},
\end{gathered}
\label{partd1}
\end{equation}
where $\Psi(\vec{x},t)=\Phi(\vec{x},s,t)\Big|_{s=\tau(\vec{x},t)}$.

Following \cite{kulq25}, we can replace the integral terms in \eqref{hartree1} with the respective moments of the function $\Psi(\vec{x},t)$. For this purpose, we introduce the modified expectations for the function $\Phi(\vec{x},s,t)$ by analogy with $\mu^{(0)}(s,t)$ and $Z(s,t)$. Let the change of variables $\vec{x}\to (s,\xi_1,...,\xi_{n-1})$ be non-degenerate. Then, we can write
\begin{equation}
\mu_{\Psi}(t,\hbar)=\dil_{\mathbb{D}}\mu_{\Phi}(s,t,\hbar)ds.
\label{moms1}
\end{equation}
We term $\mu_{\Phi}(s,t,\hbar)$ as the zeroth-order moment of the function $\Phi(\vec{x},s,t,\hbar)$. The $m$-th order central moments are defined as follows:
\begin{equation}
\begin{gathered}
\dac{1}{m!}\sum_{\sigma\in S_m}\dil_{{\mathbb{R}}^n}\Big(\Phi^{*}(\vec{x},s,t,\hbar)\Delta \hat{z}_{j_{\sigma(1)}} \Delta\hat{z}_{j_{\sigma(2)}}... \Delta\hat{z}_{j_{\sigma(m)}} \Phi(\vec{x},s,t,\hbar)\Big)\Big|_{s=\tau(\vec{x},t)}d\vec{x} =\\
=\dil_{\mathbb{D}}\Delta_{\Phi,j_1 j_2 ... j_m}(s,t,\hbar)ds.
\end{gathered}
\label{moms1}
\end{equation}
Here, $S_m$ is the symmetric group \cite{jacobson09,obukhov2024}, i.e. the operator in the left-hand size of the equation \eqref{moms1} is Weyl-ordered. We will also use the contracted notation $\Delta_{\Phi,j_1 j_2 ... j_m}=\llan \Delta \hat{z}_{(j_1} \Delta\hat{z}_{j_2}... \Delta\hat{z}_{j_m)}\rran$. In particular, $\mu_{\Phi}(s,t,\hbar)=\llan 1 \rran$. Note that we assume $\mu_{\Phi}(s,t,\hbar)$ and $\Delta_{\Phi,j_1 j_2 ... j_m}(s,t,\hbar)$ being real-valued functions. Such assumption does not lead to loss of generality since the value of the left-hand sided integral in \eqref{moms1} is real.

The asymptotic expansion of the operator $\hat{L}[\Phi]$ reads
\begin{equation}
\hat{L}=\hat{L}^{(0)}+\hat{L}^{(1)}+\hat{L}^{(2)}+...,
\label{lopexp1}
\end{equation}
where $\hat{L}^{(m)}=\hat{\Or}(\hbar^{m/2})$, $m=1,2,...$. In view of \eqref{pros17}, we can put $\hat{L}^{(0)}\equiv 0$ in our specific expansion. The second term can be put as follows:
\begin{equation}
\hat{L}^{(1)}=\langle \vec{P}_s(s,t),\Delta\vec{x}\rangle - \langle \vec{X}_s(s,t),\Delta\hat{\vec{p}}\rangle=-\langle Z_s(s,t),J\Delta\hat{z}\rangle.
\label{lopexp2}
\end{equation}
Let the asymptotic expansion of $\Phi(\vec{x},s,t,\hbar)$ read
\begin{equation}
\begin{gathered}
\Phi(\vec{x},s,t,\hbar)=\Phi^{(0)}(\vec{x},s,t,\hbar)+\hbar^{1/2}\Phi^{(1)}(\vec{x},s,t,\hbar)+\hbar^{1}\Phi^{(2)}(\vec{x},s,t,\hbar)+...,\\
\Phi^{(m)}(\vec{x},s,t,\hbar)\in {\mathcal{P}}_\hbar^{s,t}(Z(s,t), S(s,t)), \quad m=1,2,...
\end{gathered}
\label{phiexp1}
\end{equation}
Then, in the power $1/2$ of $\hbar$, we have the following equation:
\begin{equation}
\hat{L}^{(1)}\Phi^{(0)}(\vec{x},s,t)=\Big(\langle \vec{P}_s(s,t),\Delta\vec{x}\rangle - \langle \vec{X}_s(s,t),\Delta\hat{\vec{p}}\rangle\Big)\Phi^{(0)}(\vec{x},s,t)=0.
\label{expeq1}
\end{equation}
The equation \eqref{expeq1} does not uniquely determines $\Phi^{(0)}(\vec{x},s,t)$. Moreover, we will show later that the condition \eqref{expeq1} is quite weak and does not put significant limitations on the function from the physical point of view. Hence, it is safe to put this condition for the whole function $\Phi(\vec{x},s,t,\hbar)$. Thus, we will seek an asymptotic solution that is subject to the condition
\begin{equation}
\hat{L}^{(1)}\Phi(\vec{x},s,t)=\Big(\langle \vec{P}_s(s,t),\Delta\vec{x}\rangle - \langle \vec{X}_s(s,t),\Delta\hat{\vec{p}}\rangle\Big)\Phi(\vec{x},s,t)=0.
\label{expeq2}
\end{equation}
for all $s$.

Under condition \eqref{expeq2}, the leading term of asymptotics, $\Phi^{(0)}(\vec{x},s,t)$, and higher approximations, $\Phi^{(m)}(\vec{x},s,t)$, $m\geq 1$, are determined by the following equations:
\begin{equation}
\begin{gathered}
\hat{L}^{(2)}\Phi^{(0)}(\vec{x},s,t)=0,\\
\hat{L}^{(2)}\Phi^{(1)}(\vec{x},s,t)=-\hbar^{-1/2}\hat{L}^{(3)}\Phi^{(0)}(\vec{x},s,t),\\
\hat{L}^{(2)}\Phi^{(2)}(\vec{x},s,t)=-\hbar^{-1}\hat{L}^{(4)}\Phi^{(0)}(\vec{x},s,t)-\hbar^{-1/2}\hat{L}^{(3)}\Phi^{(1)}(\vec{x},s,t),\\
...
\end{gathered}
\label{expeq3}
\end{equation}

The operator $\hat{L}^{(2)}$ plays a crucial role in the construction of asymptotic solutions. Its explicit form reads as follows under condition \eqref{expeq2}:
\begin{equation}
\begin{gathered}
\hat{L}^{(2)}=-i\hbar\pa_t+ V(Z(s,t),t)+ \varkappa \dil_{{\mathbb{D}}}dr\, \mu_{\Phi}(r,t,\hbar)W(Z(s,t),Z(r,t),t)+\\
+\varkappa \dil_{{\mathbb{D}}}dr\, \Big(W_{z_i}(Z(s,t),Z(r,t),t)\Delta_{\Phi,i}(r,t,\hbar)+W_{z_i z_j}(Z(s,t),Z(r,t),t)\Delta_{\Phi,ij}(r,t,\hbar)\Big)+\\
+\varkappa \dil_{{\mathbb{D}}}dr\, W_{p_k}(Z(s,t),Z(r,t),t)\dac{X_{r,k}(r,t)}{\vec{X}_{r}^2(r,t)}\Pi_{\Phi}(r,t,\hbar)+\\
+\Big(V_{z_i}(Z(s,t),t)+ \varkappa \dil_{{\mathbb{D}}}dr\, \mu_{\Phi}(r,t,\hbar) W_{z_i}(Z(s,t),Z(r,t),t)\Big)\Delta \hat{z}_i+\\
+\Delta\hat{z}_i \Big(V_{z_i z_j}(Z(s,t),t)+ \varkappa \dil_{{\mathbb{D}}}dr\, \mu_{\Phi}(r,t,\hbar)W_{z_i z_j}(Z(s,t),Z(r,t),t)\Big)\Delta \hat{z}_j-\\
-i\hbar \Lambda \bigg(\breve{V}(Z(s,t),t)+ \varkappa \dil_{{\mathbb{D}}}dr\, \mu_{\Phi}(r,t,\hbar)\breve{W}(Z(s,t),Z(r,t),t)\bigg).
\end{gathered}
\label{expeq4}
\end{equation}
Here, the summation over repeated indices $i,j=1,...,2n$ and $k=1,...,n$ is implied. The function $\Pi_{\Phi}(s,t,\hbar)$ is the real-valued function given by
\begin{equation}
\dil_{{\mathbb{R}}^n}\Big(\Phi^{*}(\vec{x},s,t,\hbar)(-i\hbar\pa_s)\Phi(\vec{x},s,t,\hbar)\Big)\Big|_{s=\tau(\vec{x},t)}d\vec{x} =\dil_{\mathbb{D}}\Pi_{\Phi}(s,t,\hbar)ds.
\label{moms1b}
\end{equation}

The equation
\begin{equation}
\hat{L}^{(2)}\Phi^{(0)}(\vec{x},s,t)=0
\label{alsq1}
\end{equation}
can be approximated with the associated linear Schr\"{o}dinger equation by analogy with \cite{shapovalov:BTS1}. Such equation is a linearized version of \eqref{alsq1} where the functions $\Delta_{\Phi,i}(s,t,\hbar)$, $\Delta_{\Phi,ij}(s,t,\hbar)$, and $\Pi_{\Phi}(s,t,\hbar)$ are obtained from the auxiliary system of equations (the so-called higher order HSE)  that are derived in Appendix \ref{app2}. In this case, the operator $\hat{L}^{(2)}$ and the solution $\Phi^{(0)}(\vec{x},s,t)$ depend on the vector $\BFC$ of integration constants for the HSE. If these integration constants are not arbitrary but are consistent with the initial condition for $\Phi(\vec{x},s,t)$, i.e. $\BFC=\BFC[\phi]$ where $\phi(\vec{x},s)=\Phi(\vec{x},s,0)$, then the respective solution $\Phi^{(0)}(\vec{x},s,t)$ of the associated linear Schr\"{o}dinger equation generates the asymptotic solution to the Cauchy problem for the original NLSE \eqref{hartree1} with the initial condition $\psi(\vec{x})=\Psi(\vec{x},0)=\phi(\vec{x},\tau(\vec{x},0))$ in the class \eqref{pth2}.

\section{Condition (\ref{expeq2})}
\label{sec3}
Let us study the condition \eqref{expeq2} in details. In the class \eqref{pth1}, this equation can be written in the following form:
\begin{equation}
\langle \vec{X}_s(s,t),\pa_{\vec{\chi}}\varphi(\vec{\chi},s,t,\hbar)\rangle=i\langle\vec{P}_s(s,t),\vec{\chi}\rangle\varphi(\vec{\chi},s,t,\hbar).
\label{acond1}
\end{equation}
Here, we have substituted the ansatz \eqref{pth1} into \eqref{expeq2} and denoted $\vec{\chi}=\dac{\Delta \vec{x}}{\sqrt{\hbar}}$.
The solution to \eqref{acond1} is given by:
\begin{equation}
\varphi\Big(\dac{\Delta\vec{x}}{\sqrt{\hbar}},s,t,\hbar\Big)=\exp\bigg(\dac{i}{2\hbar}\dac{\big(\langle \vec{P}_s(s,t),\Delta\vec{x}\rangle\big)^2}{\langle \vec{P}_s(s,t),\vec{X}_s(s,t)\rangle}\bigg)\tilde{\varphi}\bigg(\Big(\Big\langle\vec{n}_1(s,t),\dac{\Delta \vec{x}}{\sqrt{\hbar}}\Big\rangle,...,\Big\langle\vec{n}_{n-1}(s,t),\dac{\Delta \vec{x}}{\sqrt{\hbar}}\Big\rangle\Big),s,t,\hbar\bigg),
\label{acond3}
\end{equation}
where $\vec{n}_1(s,t)$, ..., $\vec{n}_{n-1}(s,t)$ are unit vectors that form basis in the $(n-1)$-dimensional space orthogonal to $\vec{X}_s(s,t)$ in ${\mathbb{R}}^n$. The function $\tilde{\varphi}$ satisfies the same conditions as $\varphi$ in the definition of the class \eqref{pth1}.
Since $X_s(s,t)$ is a tangent vector to the curve $\vec{x}=\vec{X}(s,t)$, $s\in{\mathbb{D}}$, the equation \eqref{acond3} shows that we can consider an arbitrary form of the wave packet in the directions orthogonal to the localization curve. On the other hand, the form of a wave packet along the curve can be varied using the dependence of $\tilde{\varphi}$ on $s$. The only limitations that is put by the condition $\eqref{pth1}$ is the "fast"\, component of the phase along the curve that is determined by the exponential factors in \eqref{acond3} and in \eqref{pth1}, i.e. by the $\vec{P}(s,t)$.

Note that the vector form \eqref{acond3} is valid only for $\langle \vec{P}_s,\vec{X}_s\rangle \neq 0$. For a case involving  $\langle \vec{P}_s,\vec{X}_s\rangle = 0$, the solution to \eqref{acond1} can be written in the coordinate form as follows:
\begin{equation}
\begin{gathered}
\varphi\Big(\dac{\Delta\vec{x}}{\sqrt{\hbar}},s,t,\hbar\Big)=\exp\bigg(\dac{i}{2\hbar}\dac{2X_{s,k}(s,t)x_k \langle \vec{P}_s(s,t),\Delta \vec{x}\rangle-x_k^2 \langle \vec{P}_s(s,t),\vec{X}_s(s,t)\rangle}{X_{s,k}^2(s,t)}\bigg)\times\\
\times \tilde{\varphi}\bigg(\Big(\Big\langle\vec{n}_1(s,t),\dac{\Delta \vec{x}}{\sqrt{\hbar}}\Big\rangle,...,\Big\langle\vec{n}_{n-1}(s,t),\dac{\Delta \vec{x}}{\sqrt{\hbar}}\Big\rangle\Big),s,t,\hbar\bigg).
\end{gathered}
\label{acond4}
\end{equation}

We require the function $\tilde{\varphi}(\xi_1,...,\xi_{n-1},s,t,\hbar)$ to be from the Schwartz space with respect to $\xi_j$, $j=1,...,n-1$. Note that points $X_{s,k}^2(s,t)=0$ are often points of removable discontinuity as in the following example.

The associated linear Schr\"{o}dinger equation, that can be written as
\begin{equation}
\hat{L}^{(2)}(s,\BFC)\Phi^{(0)}(\vec{x},s,t)=0,
\label{acond5}
\end{equation}
can be solved explicitly for any fixed $\BFC$ and $s$ since it is quadratic in $\vec{x}$ and $\hat{\vec{p}}$. Its solution generate the asymptotic solutions to the Cauchy problem \eqref{hartree1} as follows:
\begin{equation}
\Psi^{(0)}(\vec{x},t)=\Big(\hat{u}\left(s,\BFC[\phi]\right)\phi(\vec{x},s)\Big)\Big|_{s=\tau(\vec{x},t)},
\label{acond6}
\end{equation}
where
\begin{equation}
\Psi^{(0)}(\vec{x},0)=\phi(\vec{x},s)\Big|_{s=\tau(\vec{x},0)}
\label{acond7}
\end{equation}
and $\hat{u}(s,\BFC)$ is linear evolution operator for the equation \eqref{acond5} parametrized by $s$ and $\BFC$ and given by
\begin{equation}
\begin{gathered}
\Phi^{(0)}(\vec{x},s,t)=\hat{u}(s,\BFC)\phi(\vec{x},s)=\dil_{{\mathbb{R}}^n}G\left(\vec{x},\vec{y},t,s,\BFC\right)\phi(\vec{y},s)d\vec{y}.
\end{gathered}
\label{green8}
\end{equation}
Here, $G\left(\vec{x},\vec{y},t,s,\BFC\right)$ is the Green function for \eqref{acond5} that is Gaussian in $\vec{x}$ and $\vec{y}$ (see, e.g., \cite{bagrov1}). Note that the relation \eqref{acond6}, which generates the asymptotic evolution operator to \eqref{hartree1}, is nonlinear with respect to the wave function since we subject $\BFC$ to the algebraic condition $\BFC=\BFC[\varphi]$.

It can be shown that the operator $\hat{L}^{(1)}(s)$ is the symmetry operator for the equation \eqref{acond5} for any $\BFC$ \cite{sym2020}. Hence, $\hat{L}^{(1)}(s)$ commutes with $\hat{u}(s,\BFC)$. This allows us to impose the condition \eqref{expeq2}, which is equivalent to \eqref{acond3} or \eqref{acond4}, only on the initial condition $\phi(\vec{x},s)=\Phi(\vec{x},s,t)\Big|_{t=0}$ and it will be automatically met for any $t>0$.

\section{Quasi-steady vortex states}
\label{sec:example}

In this Section, we deal with the quasi-steady vortex states corresponding to the solutions of the NLSE \eqref{hartree1}, which have the semiclassical counterpart as it will be shown. We will consider it within the framework of our geometric approach.

The problem under consideration is treated using the following two-dimensional ($n=2$) model:
\begin{equation}
\begin{gathered}
\bigg\{\dac{-i\hbar\pa_t}{1-i\hbar\Lambda} +\hat{\vec{p}}\,^2+\langle\vec{x},K\vec{x}\rangle + k_4|\vec{x}|^4 -\omega (\hat{p}_1 x_2 - \hat{p}_2 x_1)+\\
+\dac{\varkappa}{\pi \gamma^2}\dil_{{\mathbb{R}}^2}\exp\bigg[-\dac{(\vec{x}-\vec{y})^2}{\gamma^2}\bigg]|\Psi(\vec{y},t)|^2 d\vec{y} \bigg\}\Psi(x,t)=0.
\end{gathered}
\label{qvs1}
\end{equation}
Here, $\omega$ is the rotation frequency of the frame, $K=k_2 {\mathbb{I}}_a$ where ${\mathbb{I}}_a$ is "almost"\, identity $2\times 2$-matrix, $k_2, k_4 \in {\mathbb{R}}$ describe the strength of the second and fourth order terms, respectively, modelling the trap potential, $\Lambda$ is the phenomenological damping strength \cite{choi98}, and $\gamma$ is the nonlocality parameter. The equation \eqref{qvs1} is the nonlocal generalization for the model of the rotating Bose--Einstein condensate \cite{physrev1}. The function $|\Psi(\vec{x},t)|^2$ has a meaning of the condensate density distribution.

The quasi-steady vortex states appear as the stage of transient processes during the formation of vortex lattices that requires the system to be open, i.e. $\Lambda\neq 0$, and the axial symmetry to be broken. In our case, we use the axially symmetric initial condition and violate the symmetry of the trap potential putting $K=k_2{\mathbb{I}}+\delta K$ where $\delta K$ is a small nonidentity diagonal matrix that determines asymmetry.

Within notations of \eqref{hartree1}, we have the following symbols for operators entering the equation \eqref{qvs1}:
\begin{equation}
\begin{gathered}
V(z,t)=\vec{p}\,^2+\langle\vec{x},K\vec{x}\rangle + k_4|\vec{x}|^4 -\omega (\hat{p}_1 x_2 - \hat{p}_2 x_1),\\
W(z,w,t)=\dac{1}{\pi \gamma^2}\exp\bigg[-\dac{(\vec{x}-\vec{y})^2}{\gamma^2}\bigg],\\
\breve{V}(z,t)=V(z,t), \qquad \breve{W}(z,w,t)=W(z,w,t).
\end{gathered}
\label{p2vprim2}
\end{equation}

Under certain conditions, the quantized vortices tends to form a dense cluster in a neighbourhood of the center of rotation \cite{physrev1}. Such behaviour reminds the semiclassical one. Then, the density minima corresponding to the phase defects of quantized vortices form one giant density dip and the condensate appear to be localized on a curve enveloping such dip. If the number of such quantized vortices $N\gg 1$ and the system admits the axial symmetry (at least, the approximate one), such curve is approximately a circle. Let us consider such case, when the initial condition is localized on a rotating circle. That yields the following initial conditions for the zeroth order HES \eqref{hes1}:
\begin{equation}
\begin{gathered}
\vec{X}(s,0)=(R \cos s, R \sin s),\\
\vec{P}(s,0)=(P_{r} \sin s, -P_{r} \cos s),\\
\mu^{(0)}(s,0)=\const=\mu_0,\\
s\in[0;2\pi].
\end{gathered}
\label{p2vprim3}
\end{equation}
Here, $R$ is the radius of the circle in the coordinate space and $P_r$ determines the rotation speed. The condition $\mu^{(0)}(s,0)=\const$ indicates that we assume the condensate being uniformly distributed along the localization circle at the initial moment of time. The quantization condition \eqref{p2hes11} for \eqref{p2vprim3} reads
\begin{equation}
\begin{gathered}
R P_{r}=N\hbar.
\end{gathered}
\label{p2vprim4}
\end{equation}
The quantity $N$ indicates the number of quantized vortices inside the boundary $\vec{x}=\vec{X}(s,t)$, $s\in[0;2\pi]$.
The abbreviated classical action \eqref{abbac1} on the curve \eqref{p2vprim3} can be written as follows:
\begin{equation}
S(s,0)=-R P_{r} s,
\label{p2vprim5}
\end{equation}
The multiplier in \eqref{acond4} reads
\begin{equation}
\begin{gathered}
\exp\bigg(\dac{i}{2\hbar}\dac{2X_{s,1}(s,0)x_1 \langle \vec{P}_s(s,0),\Delta \vec{x}\rangle-x_1^2 \langle \vec{P}_s(s,0),\vec{X}_s(s,0)\rangle}{X_{s,1}^2(s,0)}\bigg)=1.
\end{gathered}
\label{p2vprim6}
\end{equation}
We see that the discontinuity in points $X_{s,1}^2(s,0)=0$ is removable and no extra phase term is added in this case.

Now we can explicitly write the initial condition $\Psi(\vec{x},0,\hbar)=\Phi(\vec{x},s,0,\hbar)\big|_{s=\tau(\vec{x},0)}$ in view of \eqref{p2vprim6}, \eqref{p2vprim5}, and \eqref{p2vprim3}. For our parametrization of the curve, we have $\tau(\vec{x},0)=\arg(x_1+i x_2)$. Let us consider a simple form of $\tilde{\varphi}(\cdot,s,t,\hbar)$ that is the Gaussian one. Then, the initial condition has the following form in the expanded space:
\begin{equation}
\begin{gathered}
\Phi(\vec{x},s,t,\hbar)\Big|_{t=0}=\sqrt{\dac{\mu_0}{\gamma R \sqrt{\pi \hbar}}}\exp\left(-\dac{(x_1 \cos s + x_2 \sin s - R)^2}{0.5\hbar}\right)\times\\ \times \exp\left(-\dac{i}{\hbar}P_{r}\left(x_1 \sin s - x_2 \cos s\right)\right)\exp\left(-\dac{i}{\hbar}R P_{r} s\right).
\end{gathered}
\label{p2vprim7}
\end{equation}
The respective initial condition in the original space, $(t,\vec{x})$, reads
\begin{equation}
\begin{gathered}
\Psi(\vec{x},t,\hbar)\Big|_{t=0}=\Phi(x,s,0,\hbar)\Big|_{s=\arg(x_1+ix_2)}=\\
=\sqrt{\dac{\mu_0}{\gamma R \sqrt{\pi \hbar}}}\exp\left(-\dac{(|\vec{x}| - R)^2}{0.5\hbar}\right)\exp\left(-\dac{i}{\hbar}R P_{\varphi} \arg(x_1+ix_2)\right).
\end{gathered}
\label{p2vprim8}
\end{equation}

Note that the initial condition \eqref{p2vprim8} is continuous under the condition \eqref{p2vprim4} but it is not differentiable in the point $x_1=x_2=0$. However, since the Gaussian function tends to zero faster than any power of $\hbar$ in this point as $\hbar\to 0$, we can readily regularize the function $\eqref{p2vprim8}$ in this point. The smooth regularization would give the yield $\Or(\hbar^{\infty})$ that does not affect the semiclassical equations. Thus, the semiclassical formalism remains valid.

\begin{figure}[!ht]
    \centering\includegraphics[width=16 cm]{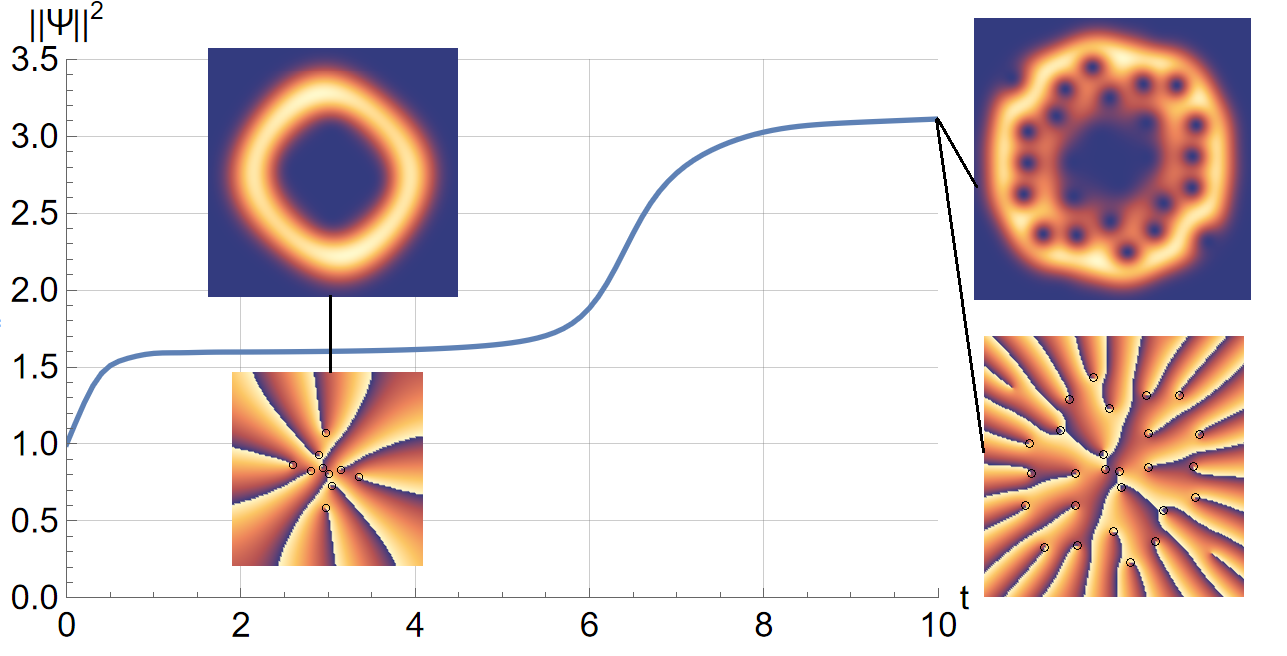} \\
  \caption{Evolution of the squared norm of the vortex state with time\label{fig1}}
\end{figure}

First, let us illustrate the dynamics that is generated by the model \eqref{qvs1} by an example of numerical solutions. The Fig. \ref{fig1} illustrates the evolution of the squared $L_2$-norm of $\Psi$ with time for $R=3$, $\Lambda=0.3$, $\varkappa=250$, $k_4=\dac{1}{16}$, $\gamma=1$, $\omega=3$, $\mu_0=\dac{1}{2\pi}$, $K=\dac{1}{4}\begin{pmatrix}1.1 & 0 \cr 0 & 0.9\end{pmatrix}$, $\hbar=1$, and $N=10$. Also, two samples of $|\Psi(\vec{x},t)|^2$ (upper) and $\arg \Psi(\vec{x},t)$ (lower) are shown for $t=3$ and $t=10$ (the color gradient reflects the density and phase, respectively, with the brighter one being the higher one). The phase defects, corresponding to the centers of quantized vortices, are circled in black for the illustrativeness.

The dynamics shown in Fig. \ref{fig1} can be conditionally divided into the following stages. During the first stage ($t\sim 0\div 0.5$), the number of particles associated with the $L_2$-norm grows rapidly until the system reaches the thermodynamical quasi-equilibrium. Note that the $L_2$-norm can either decrease for other initial condition. Since the asymmetry is small, the localization curve is still almost the ideal circle. The second stage ($t\sim 1\div 5$) is the quasi steady vortex state. Here, the number of particles is almost constant and the localization circle is slowly deforming. When the deformation reaches the critical degree, the new transient process (next stage) starts that leads to the formation of new vortex lattice with a different number of quantized vortices.

\begin{figure}[!ht]
    \centering\includegraphics[width=16 cm]{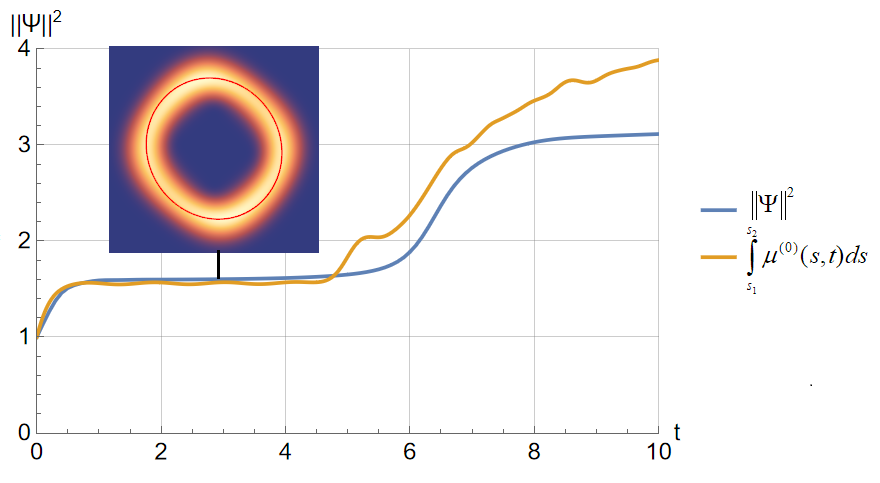} \\
  \caption{Comparison of semiclassical description with numerical solutions to \eqref{qvs1} \label{fig2}}
\end{figure}

Now, let us compare the dynamics in Fig. \ref{fig1} with the one predicted with a semiclassical approximation given by the solutions to \eqref{hes1}. The comparison is given in Fig. \ref{fig2}. The $L_2$-norm is approximated by $\dil_{s_1}^{s_2} \mu^{(0)}(s,t)ds$. The red curve stands for $\vec{x}=\vec{X}(s,3)$, $s\in[0;2\pi]$.

\begin{figure}[!ht]
    \centering\includegraphics[width=13 cm]{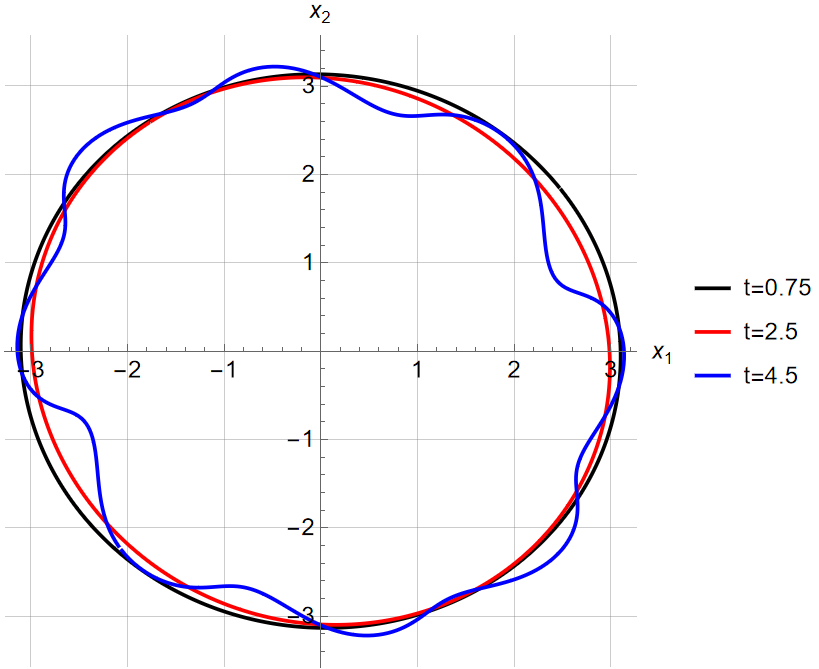} \\
  \caption{Evolution of the curve $\vec{x}=\vec{X}(s,t)$ \label{fig3}}
\end{figure}

Evidently, the semiclassical approximation is in a good agreement with the direct numerical solutions for the first two stages. In particular, it looks valid for the description of quasi steady vortex states. Let us consider them in details within the semiclassical geometric approach. As said above, the evolution of the quasi-steady vortex state is associated with the deformation of the localization curve, specifically $\vec{x}=\vec{X}(s,t)$. In Fig. \ref{fig3}, the evolution of this curve is shown. We start from the ideal circle that is hardly deformed for the early phase of the quasi steady vortex state ($t=0.75$).

For the axially symmetric problem, the circle would never lose its symmetry, i.e. the deformation is the results of the small asymmetry. Hence, we can linearize the system \eqref{hes1} for small $\delta K$. Let the localization curve be the ideal rotating circle in the beginning of the quasi steady vortex state. We denote this moment of time as $t=t_0$. Then, nondeformed circle $\vec{x}=\vec{\overline{X}}(s,t)$, $s\in[0;2\pi]$, is given by
\begin{equation}
\begin{gathered}
\vec{\overline{X}}(s,t)=\left(\overline{R}\cos(\overline{\omega}t+s+s_0),\overline{R}\sin(\overline{\omega}t+s+s_0)\right),\\
\vec{\overline{P}}(s,t)=\left(\overline{P}_r\sin(\overline{\omega}t+s+s_0),-\overline{P}_r\cos(\overline{\omega}t+s+s_0)\right),\\
\overline{\mu}(s,t)=\const=\overline{\mu_0}.
\end{gathered}
\label{p2vprim10}
\end{equation}
The functions \eqref{p2vprim10} are the zeroth order approximation to the localization curve with respect to the small deformation $\delta K$. The value $s_0$ is determined by the first stage of the transient process before $t=t_0$ and we will be put $s_0=0$ thereinafter without loss of generality. The functions \eqref{p2vprim10} correspond to the solution of \eqref{hes1} that yields the curve invariant with respect to the time shift for $\delta K=0$. Then, if we put $\delta K=0$, the values of $\overline{\mu_0}$, $\overline{\omega}$, $\overline{R}$, and $\overline{P}_r$ can readily be obtained from \eqref{hes1} as follows:
\begin{equation}
\begin{gathered}
\overline{R}(\overline{\omega}-\omega)+2\overline{P}_r=0,\\
\overline{P}_r(\overline{\omega}-\omega)+2\overline{R}\left(k_2+2k_4 \overline{R}^2\right)-\dac{\varkappa \overline{\mu}_0}{\gamma^4}\overline{R}\exp\left(-\dac{2\overline{R}^2}{\gamma^2}\right)\left(I_0\left(\dac{2\overline{R}^2}{\gamma^2}\right)-I_1\left(\dac{2\overline{R}^2}{\gamma^2}\right)\right)=0,\\
\overline{R}^2\left(1+k_2+k_4\overline{R}^2\right)-\omega\overline{R}+\dac{2\varkappa\overline{\mu}_0}{\gamma^2}I_0\left(\dac{2\overline{R}^2}{\gamma^2}\right)=0.
\end{gathered}
\label{p2deform1}
\end{equation}
Here, $I_{n}(z)$ are the modified Bessel functions of the first kind. The system \eqref{p2deform1} determines $\overline{\mu_0}$, $\overline{\omega}$, $\overline{R}$, and $\overline{P}_r$ under the quantization condition \eqref{p2vprim4}. The equation \eqref{p2deform1} are evidently nonlinear.
Let us present the localization curve during the quasi steady state mode as follows:
\begin{equation}
\begin{gathered}
\vec{X}(s,t)=\vec{\overline{X}}(s,t)+\delta\vec{X}(s,t)+\Or(d^2),\\
\vec{P}(s,t)=\vec{\overline{P}}(s,t)+\delta\vec{P}(s,t)+\Or(d^2),\\
\mu^{(0)}(s,t)=\overline{\mu}_0+\delta\mu(s,t)+\Or(d^2),
\end{gathered}
\label{p2deform2}
\end{equation}
where $d=||\delta K||$ is a norm of the matrix $\delta K$ (small parameter). The functions $\delta Z(s,t)=\left(\delta\vec{P}(s,t),\delta\vec{X}(s,t)\right)$ and $\delta\mu(s,t)$ determine the deformation of the localization curve in the linear approximation with respect to $\delta K$. We assume the initial deformation for $t=t_0$ being negligible, i.e. $\delta Z(s,t_0)=0$, $\delta\mu(s,t_0)=0$. Then, the functions $\delta Z(s,t)=\left(\delta\vec{P}(s,t),\delta\vec{X}(s,t)\right)$ and $\delta\mu(s,t)$ are solutions to the following Cauchy problem:
\begin{equation}
\begin{gathered}
\dac{\pa \delta Z(s,t)}{\pa t}=A(\overline{\omega}t+s)\delta Z(s,t)+\\
+\dil_{0}^{2\pi}\left(B(\overline{\omega}t+s,\overline{\omega}t+r)\delta Z(r,t)+C(\overline{\omega}t+s,\overline{\omega}t+r)\delta\mu(r,t)\right)dr-2 \begin{pmatrix} \delta K & 0 \cr 0 & 0\end{pmatrix},\\
\dac{\pa \delta \mu(s,t)}{\pa t}=2\Lambda\overline{\mu}_0\dil_{0}^{2\pi}\left(\overline{\mu}_0\langle C(\overline{\omega}t+s,\overline{\omega}t+r),\delta Z(r,t)\rangle-\dac{\delta \mu(r,t)}{\gamma^2 \pi}\exp\left(\dac{2\overline{R}^2\left(\cos(s-r)-1\right)}{\gamma^2}\right)\right) dr+\\
  +2\Lambda\overline{\mu}_0 \langle \delta Z(s,t),J\dot{Z}^{(0)}(s,t)\rangle -2\Lambda\overline{\mu}_0 \langle \vec{X}^{(0)}(s,t),\delta K\cdot\vec{X}^{(0)}(s,t)\rangle,\\
\end{gathered}
\label{p2deform3}
\end{equation}
\begin{equation}
\begin{gathered}
A(s)=\begin{pmatrix} 0 & \omega & a+b\cos 2s & b \sin 2s \cr -\omega & 0 & b \sin 2s & a-b\cos 2s \cr 2 & 0 & 0 & \omega \cr 0 & 2 & -\omega & 0 \end{pmatrix},\\
a=-2k_2-8k_4\overline{R}^2+
\dac{4\varkappa\overline{\mu}_0}{\gamma^6}\exp\left(-\dac{2\overline{R}^2}{\gamma^2}\right)\left(2\overline{R}^2I_1\left(\dac{2\overline{R}^2}{\gamma^2}\right)+
\left(\gamma^2-2\overline{R}^2\right){}_0F_1\left(;1;\dac{\overline{R}^4}{\gamma^4}\right)\right), \\
b=-4k_4 \overline{R}^2+
\dac{4\varkappa\overline{\mu}_0}{\gamma^6}\exp\left(-\dac{2\overline{R}^2}{\gamma^2}\right)\left(\left(\gamma^2+2\overline{R}^2\right)I_1\left(\dac{2\overline{R}^2}{\gamma^2}\right)-
2\overline{R}^2{}_0F_1\left(;1;\dac{\overline{R}^4}{\gamma^4}\right)\right),\\
\end{gathered}
\label{p2deform4}
\end{equation}
\begin{equation}
\begin{gathered}
B(s,r)=\begin{pmatrix} 0 & 0 \cr 0 & \tilde{B}(s,r) \end{pmatrix},\\
\tilde{B}(s,r)=\begin{pmatrix} \tilde{a}(s-r)+\tilde{b}(s-r)\cos(s+r) & \tilde{b}(s-r)\sin(s+r) \cr \tilde{b}(s-r)\sin(s+r) & \tilde{a}(s-r)-\tilde{b}(s-r)\cos(s+r) \end{pmatrix},\\
\tilde{a}(s)=\dac{2\varkappa\overline{\mu}_0}{\gamma^6\pi}\exp\left(-\dac{2\overline{R}^2(1-\cos s)}{\gamma^2}\right)\left(\gamma^2-4\overline{R}^2\sin^2\!\left(\dac{s}{2}\right)\right),\\
\tilde{b}(s)=\dac{8\varkappa\overline{\mu}_0\overline{R}^2}{\gamma^6\pi} \exp\left(-\dac{2\overline{R}^2(1-\cos s)}{\gamma^2}\right)\sin^2\!\left(\dac{s}{2}\right),\\
C(s,r)=\dac{4\varkappa \overline{R}}{\gamma^4 \pi}\exp\left(\dac{2\overline{R}^2\left(\cos(s-r)-1\right)}{\gamma^2}\right)\sin\left(\dac{r-s}{2}\right)\cdot\begin{pmatrix} -\sin\left(s+r\right) \cr \cos\left(s+r\right) \cr 0 \cr 0 \end{pmatrix}
\end{gathered}
\label{p2deform4b}
\end{equation}
where $\begin{pmatrix} \delta K & 0 \cr 0 & 0\end{pmatrix}$ and $B(s,r)$ are $4\times 4$ block matrices, and ${}_0F_1(;\cdot;\cdot)$ is the confluent hypergeometric function \cite{beitmen1}.

The equations \eqref{p2deform3} show that the weight function $\mu(s,t)$ changes with time during the quasi steady vortex state despite the fact that $\dil_{0}^{2\pi}\mu(s,t)ds$ (total number of particles) is almost constant, i.e. the distribution of the matter along the curve changes with the deformation. It explains the fact that the transformation of the vortex lattices can only occur in nonconservative (open) systems where $\mu(s,t)$ is not conserved.

Since the deformation of the localization curve from its initial circle-shaped form is associated with the evolution of quasi steady vortex state, it is of interest to give the formal criterion for the end of such state and the beginning of the transition to a new vortex state. Thus, we propose the critical index for it that is the convexity of the localization curve. When it becomes essentially non-convex, it can be treated as the beginning of the transition to a new vortex state with a different number of quantized vortices. Let us clarify the nature of such assumption. As we mentioned above, the value $N$ is conserved quantity for \eqref{hes1}. On the other hand, it can be treated as the number of inner phase defects. The transition to a vortex state with a different number of quantized vortices is a process when the outer phase defects become the inner one or vice versa. Here, the "outer"\, phase defects are interpreted as the ones that are located in the area where the density of the matter is negligibly small. When the localization curve, which is purely semiclassical concept, becomes essentially non-convex, we cannot more treat its inner domain as the domain where all of the inner phase defects of continuous matter are localized. Such reasoning also naturally leads us to the fact that the semiclassical approximation cannot give the accurate description of the transient process after the second stage that was already shown in Fig. \ref{fig2}.

\section{Conclusion}
\label{sec:con}

We have proposed the approach to constructing the asymptotic solutions to the Cauchy problem for the nonlocal NLSE with an anti-Hermitian terms. Our solutions are semiclassically localized on the one-dimensional manifold (the curve) that evolves with time. Our formalism is based on the Maslov complex germ method. We lift the dimension of the original NLSE by introducing additional independent variable and modify the class of functions where the semiclassical solutions are sought. This allows us to apply the theory of semiclassically localized states to the problem under consideration. In this sense, the approach has some similarities with the parabolic deformation of Riemannian manifolds proposed by Hamilton \cite{hamilton82} that opened up new perspectives in the study of the global properies of Riemannian manifolds \cite{perelman2002}, gravitation and field theory \cite{cao2003,topping2006,odintsov2012,obukhov2023}, and even reaction-diffusion systems \cite{ivancevic11}. In our approach, some global properties of the asymptotic solution are dictated by the geometry of the one-dimensional localization manifold $\Lambda_t$.

The solutions obtained can describe the behavior of the open quantum systems with nontrivial geometry within the semiclassical approximation. We demonstrate it with the specific model NLSE that describes the evolution of a vortex state in the rotating BEC. We show that such evolution can have the stage that is treated as the quasi-steady vortex state. This state has the semiclassical counterpart and is associated with the slowly deforming circle-like curve. Within such geometrical interpretation, the linearized equations are derived that describe such deformation. Also, we propose that the curve convexity is the critical index that is associated with the lifetime of the quasi-steady vortex state.

Since our approach naturally deals with the nonlocal models, it is of interest to study its applicability to the models with the nonlocal linear terms expressed via the fractional kinetic terms that account for the nonlocal properties of the media \cite{laskin2006,tarasov11}. These are the prospects of this work.

%

\appendix

\section{Pseudo-differential operators}
\label{app0}
Let functions $A(\vec{p},\vec{x},t,\hbar)$, $\vec{p},\vec{x}\in{\mathbb{R}}^{n}$ satisfy the following conditions for every fixed $t\geq 0$:\\
1) $A(\vec{p},\vec{x},t,\hbar)\in C^{\infty}$ with respect to $\vec{p}$ and $\vec{x}$;\\
2) $A(\vec{p},\vec{x},t,\hbar)$ and all its derivatives grow not faster than polynomials of $|\vec{p}|$ and $|\vec{x}|$ as $|\vec{p}|,|\vec{x}|\to \infty$;\\
3) $A(\vec{p},\vec{x},t,\hbar)$ regularly depends on the parameter $\hbar$ in a neighborhood of $\hbar=0$. \\
These functions form the class ${\mathcal{S}}_\hbar^t$.
We also use the brief notation $A(z,t,\hbar)=A(\vec{p},\vec{x},t,\hbar)$, $z=(\vec{p},\vec{x})\in{\mathbb{R}}^{2n}$.

\begin{defin}
A pseudo-differential Weyl-ordered operator is an operator $\hat{A}=A(\hat{z},t,\hbar)=A(\hat{\vec{p}},\vec{x},t,\hbar)$ that is defined by \cite{Maslov1}
\begin{equation}
A(\hat{\vec{p}},\vec{x},t,\hbar)\Phi(\vec{x},t,\hbar)=\dac{1}{(2\pi\hbar)^n}\dil_{{\mathbb{R}}^{2n}}d\vec{p}d\vec{y} \exp\Big(\dac{i}{\hbar}\langle \vec{p},\vec{x}-\vec{y}\rangle\Big)A\Big(\vec{p},\dac{\vec{x}+\vec{y}}{2},t,\hbar\Big)\Phi(\vec{y},t,\hbar),
\label{pseud1}
\end{equation}
where $A(\vec{p},\vec{x},t,\hbar)\in{\mathcal{S}}_\hbar^t$ and $\Psi(\vec{x},t,\hbar)\in{\mathbb{S}}$ for fixed $t$, $\hbar$. Here, ${\mathbb{S}}$ is the Schwartz space, and $\langle a, b \rangle=\sum_{i=1}^{n}a_i b_i$ is the Euclidian scalar product.

The function $A(z,t,\hbar)$ in \eqref{pseud1} is termed the Weyl symbol of the operator $\hat{A}=A(\hat{z},t,\hbar)$. We denote by ${\mathcal{A}}_\hbar^t$ the set of pseudo-differential operators defined above.
\end{defin}

Note that the Weyl ordering of a linear operator is "symmetric"\, since it leads to the conventional symmetrization for differential operators. For example, if $A(p,x,t,\hbar)=px$, $n=1$, the respective Weyl-ordered operator reads $A(\hat{p},x,t,\hbar)=\dac{1}{2}\left(\hat{p}x+x\hat{p}\right)$.

For the semiclassical approximation theory, the following properties of pseudo-differential operators are useful. Let the pseudo-differential operators $C(\hat{z},t)$ and $D(\hat{z},t)$ be given by
\begin{equation}
\begin{gathered}
C(\hat{z},t)=\big[A(\hat{z},t),B(\hat{z},t)\big]=A(\hat{z},t)B(\hat{z},t)-B(\hat{z},t)A(\hat{z},t), \\ D(\hat{z},t)=\big[A(\hat{z},t),B(\hat{z},t))\big]_{+}=A(\hat{z},t)B(\hat{z},t)+B(\hat{z},t)A(\hat{z},t),
\end{gathered}
\label{limpo0}
\end{equation}
where $A(\hat{z},t)$ and $B(\hat{z},t)$ are pseudo-differential operators with the Weyl symbols $A(z,t)$ and $B(z,t)$ respectively.

Then, their Weyl symbols $C(z,t)$ and $D(z,t)$ obey the following relations \cite{multiind}:
\begin{equation}
\begin{gathered}
\lim_{\hbar\to 0} \dac{C(z,t)}{i\hbar}=\big\{A(z,t),B(z,t)\big\}, \qquad \lim_{\hbar\to 0} D(z,t)=2A(z,t)B(z,t).
\end{gathered}
\label{limpo1a}
\end{equation}
where $\big\{A(z,t),B(z,t)\big\}=\bigg\langle\displaystyle\frac{\partial A(z,t)}{\partial z},J \frac{\partial B(z,t)}{\partial z}\bigg\rangle$ is the Poisson bracket, $J=\begin{pmatrix} 0 & -{\mathbb{I}}_{n\times n} \cr {\mathbb{I}}_{n\times n} & 0 \end{pmatrix}$.

\section{Derivation of the zeroth order HES \eqref{hes1}}
\label{app0b}
Let us derive the equation for $\dot{\mu}_{\Psi}(t,\hbar)$ by the substitution of $\pa_t\Psi(\vec{x},t)$ from \eqref{hartree1} into $\dot{\mu}_{\Psi}(t,\hbar)$:
\begin{equation}
\begin{gathered}
\dot{\mu}_{\Psi}(t,\hbar)=-2\Lambda \displaystyle\int\limits_{{\mathbb{R}}^n}
d\vec{x}\,\Psi^{*}(\vec{x},t;\hbar) \breve{H}(\hat{z},t)[\Psi]\Psi(\vec{x},t;\hbar)=
-2\Lambda\langle\Psi|\breve{H}[\Psi]|\Psi \rangle=\\
=-2\Lambda\left(\langle\Psi|\breve{V}(\hat{z},t)|\Psi \rangle+\varkappa \langle\Psi|\dil_{{\mathbb{R}}^n}d\vec{y}\,\Psi^{*}(\vec{y},t)\breve{W}(\hat{z},\hat{w},t)\Psi(\vec{y},t)|\Psi \rangle\right).
\end{gathered}
 \label{sig2}
 \end{equation}

Under the condition \eqref{def1a}, we have
\begin{equation}
\begin{gathered}
\lim\limits_{\hbar\to 0}\langle\Psi|\breve{V}(\hat{z},t)|\Psi \rangle=\dil_{{\mathbb{D}}}ds\,\mu^{(0)}(s,t)\breve{V}(Z(s,t),t),\\
\lim\limits_{\hbar\to 0}\langle\Psi|\dil_{{\mathbb{R}}^n}d\vec{y}\,\Psi^{*}(\vec{y},t)\breve{W}(\hat{z},\hat{w},t)\Psi(\vec{y},t)|\Psi \rangle=\dil_{{\mathbb{D}}}ds\,\mu^{(0)}(s,t)\dil_{{\mathbb{D}}}dr\,\mu^{(0)}(r,t)\breve{W}(Z(s,t),Z(r,t),t).
\end{gathered}
\label{limop1}
\end{equation}
Then, we have the following form of the equation \eqref{sig2} in the limit $\hbar\to 0$:
\begin{equation}
\dil_{{\mathbb{D}}}ds\,\dot{\mu}^{(0)}(s,t)=-2\Lambda\dil_{{\mathbb{D}}}ds\, \mu^{(0)}(s,t)\left( \breve{V}(Z(s,t),t)+\varkappa \dil_{{\mathbb{D}}}dr\,\mu^{(0)}(r,t)\breve{W}(Z(s,t),Z(r,t),t)\right).
\label{limsig2}
\end{equation}

In a similar manner, we obtain the exact equation for $A_{\Psi}(t)$ where $A_{\Psi}(t)=\langle \hat{A} \rangle_{\Psi}=\langle \Psi | \hat{A} | \Psi \rangle$:
\begin{align}
&\displaystyle\frac{\partial}{\partial t}\langle\hat{A}(t)\rangle_\Psi=\bigg\langle\displaystyle\frac{\partial \hat{A}(t)}{\partial t} \bigg\rangle_\Psi +\frac{i}{\hbar}\big\langle \big[H(\hat{z},t)[\Psi],\hat{A}(t)\big]
 \big\rangle_\Psi-\Lambda \big\langle \big[\breve{H}(\hat{z},t)[\Psi],\hat{A}(t)\big]_{+}
 \big\rangle_\Psi=\cr
 &= \bigg\langle\displaystyle\frac{\partial \hat{A}(t)}{\partial t} \bigg\rangle_\Psi+\dac{i}{\hbar} \big\langle [V(\hat{z},t),A(\hat{z},t)]\big\rangle_\Psi - \Lambda\big\langle [\breve{V}(\hat{z},t),A(\hat{z},t)]_{+}\big\rangle_\Psi + \cr
&+\varkappa \bigg\langle\dil_{{\mathbb{R}}^n}d\vec{y}\, \Psi^{*} \Big(\dac{i}{\hbar}[W(\hat{z},\hat{w},t), A(\hat{z},t)]-\Lambda [\breve{W}(\hat{z},\hat{w},t), A(\hat{z},t)]_{+} \Big)\Psi(\vec{y},t)\bigg\rangle_\Psi.
  \label{mean2}
\end{align}

The properties \eqref{limpo1a} of pseudo-differential operators yield the following equation for the operator $\hat{A}=A(\hat{z},t)$ with a Weyl symbol $A(z,t)$ in the limit $\hbar\to 0$:
\begin{equation}
\begin{gathered}
\dac{d}{dt}\dil_{{\mathbb{D}}}ds\,\mu^{(0)}(s,t)A(Z(s,t),t)=\dil_{{\mathbb{D}}}ds\,\mu^{(0)}(s,t)\Bigg(\dac{\pa A(z_s,t)}{\pa t}-\left\{V(z_s,t),A(z_s,t)\right\}-2\Lambda \breve{V}(z_s,t)A(z_s,t)+\\
+\varkappa\dil_{{\mathbb{D}}}dr\, \mu^{(0)}(r,t)\Big(-\left\{W(z_s,w_r,t),A(z_s,t)\right\}-2\Lambda\breve{W}(z_s,w_r,t)A(z_s,t)\Big)\Bigg)\Big|_{z_s=Z(s,t),\,w_r=Z(r,t)}.
\end{gathered}
\label{limop2}
\end{equation}
In particular, for $A(z,t)=z$, we have
\begin{equation}
\begin{gathered}
\dac{d}{dt}\dil_{{\mathbb{D}}}ds\,\mu^{(0)}(s,t)Z(s,t)=\dil_{{\mathbb{D}}}ds\,\mu^{(0)}(s,t)\Bigg(JV_z(Z(s,t),t)-2\Lambda \breve{V}(Z(s,t),t)Z(s,t)+\\
+\varkappa\dil_{{\mathbb{D}}}dr\, \mu^{(0)}(r,t)\Big(J W_z(Z(s,t),Z(r,t),t)-2\Lambda\breve{W}(Z(s,t),Z(r,t),t)Z(s,t)\Big)\Bigg).
\end{gathered}
\label{limop3}
\end{equation}

The system \eqref{limsig2}, \eqref{limop3} is the weak form of the following system of $(2n+1)$ integro-differential equations:
\begin{equation}
\begin{gathered}
\dot{\mu}^{(0)}(s,t)=-2\Lambda\mu^{(0)}(s,t)\left( \breve{V}(Z(s,t),t)+\varkappa \dil_{{\mathbb{D}}}dr\,\mu^{(0)}(r,t)\breve{W}(Z(s,t),Z(r,t),t)\right), \\
\dot{Z}(s,t)=JV_z(Z(s,t),t)+\varkappa\dil_{{\mathbb{D}}}dr\, \mu^{(0)}(r,t)J W_z(Z(s,t),Z(r,t)).
\end{gathered}
\label{hes11}
\end{equation}

\section{Proof of the Statement \ref{statement1}}
\label{app0c}

The first two equations of the system \eqref{hes1} can be written as
\begin{equation}
\begin{gathered}
\dot{\vec{P}}(s,t)=-H_{\vec{x}}\left(Z(s,t),t\right),\\
\dot{\vec{X}}(s,t)=H_{\vec{p}}\left(Z(s,t),t\right),
\end{gathered}
\label{bzpr1}
\end{equation}
where $H\left(Z(s,t),t\right)$ is real-valued infinitely smooth function, $H_{z}\left(Z(s,t),t\right)=\dac{H(z,t)}{\pa z}\Big|_{z=Z(s,t)}$, $z=(\vec{p},\vec{x})$.

Let us differentiate \eqref{p2hes11} with respect to $t$, apply the integration by parts:
\begin{equation}
\begin{gathered}
\pa_t F(t)=\dil_{s_1}^{s_2} \left(\langle \dot{\vec{P}}(s,t),\vec{X}_s(s,t)\rangle+\langle \vec{P}(s,t),\dot{\vec{X}}_s(s,t)\rangle\right) ds=\langle \dot{\vec{P}}(s,t),\vec{X}(s,t)\rangle\Big|_{s_1}^{s_2}-\\
-\dil_{s_1}^{s_2} \left(\langle \dot{\vec{P}}_s(s,t),\vec{X}(s,t)\rangle-\langle \vec{P}(s,t),\dot{\vec{X}}_s(s,t)\rangle\right) ds=\\
=\dil_{s_1}^{s_2} \left(\langle \vec{P}(s,t),\dot{\vec{X}}_s(s,t)\rangle-\langle \dot{\vec{P}}_s(s,t),\vec{X}(s,t)\rangle\right) ds.
\end{gathered}
\label{bzpr2}
\end{equation}
Here, we have taken into account that $\Lambda_t$ is a closed curve. Now, we substitute \eqref{bzpr1} into \eqref{bzpr2} and obtain the following:
\begin{equation}
\begin{gathered}
\pa_t F(t)=\dil_{s_1}^{s_2} \left(\langle \vec{P}(s,t),\pa_s H_{\vec{p}}(Z(s,t),t)\rangle+\langle \pa_s H_{\vec{x}}(s,t),\vec{X}(s,t)\rangle\right) ds=\\
=\dil_{s_1}^{s_2} \left(\langle \vec{P}(s,t),H_{\vec{p}z}(Z(s,t),t)Z_s(s,t)\rangle+\langle H_{\vec{x}z}(s,t)Z_s(s,t),\vec{X}(s,t)\rangle\right) ds=\\
=\dil_{s_1}^{s_2} \langle Z(s,t),H_{zz}(Z(s,t),t)Z_s(s,t)\rangle ds=\oint\limits_{\Lambda_t} \langle z^T H_{zz}(z,t), dz\rangle.
\end{gathered}
\label{bzpr3}
\end{equation}
One readily gets that $\nabla_z \times \left(z^T H_{zz}(z,t)\right)=0$, i.e. the integral \eqref{bzpr3} is equal to zero due to Stokes' theorem.

\section{The second order HES}
\label{app2}

Here, we derive the second order HES that allows one to linearize the NLSE in the expanded space.

The Weyl-ordered operator $H(\hat{z},t)$ can be presented as follows in the expanded space:
\begin{equation}
\begin{gathered}
H(\hat{\vec{p}},\vec{x},t)f(\vec{x},s,t)=H(\hat{\vec{\pi}}-i\hbar\tau_{\vec{x}}(\vec{x},t)\pa_s,\vec{x},t)f(\vec{x},s,t)=\\
=H\Big(\big(\hat{\vec{\pi}}-i\hbar\tau_{\vec{x}}(\vec{x},t)\pa_s-\vec{P}(r,t)\big)+\vec{P}(r,t),\big(\vec{x}-\vec{X}(r,t)\big)+\vec{X}(r,t),t\Big)f(\vec{x},s,t)=\\
=\sum_{i,j=0}^{\infty}\dac{1}{m!q!}H_{p_{i_1}...p_{i_m}x_{j_1}...x_{j_q}}\Big(\vec{P}(r,t),\vec{X}(r,t),t\Big)\cdot\\
\cdot\bigg\{\Big(\hat{\vec{\pi}}-i\hbar\tau_{\vec{x}}(\vec{x},t)\pa_s-\vec{P}(r,t),\vec{x}-\vec{X}(r,t)\Big)^{(i_1,...,i_m,j_1,...,j_q)}\bigg\}f(\vec{x},s,t).
\end{gathered}
\label{razras0}
\end{equation}
where $\{\cdot\}$ is the Weyl symmetrization. If we substitute $r=s$ into \eqref{razras0}, the right-hand side of \eqref{razras0} becomes the asymptotic series due to the estimates \eqref{pros16}, \eqref{pros17}. Thus, the direct calculation of the expansion \eqref{razras0} and the subsequent substitution of $r=s$ yields us the following estimate:
\begin{equation}
\begin{gathered}
H(\hat{z})=H\big(Z(s)\big)+H_{z_a}\big(Z(s)\big) \Delta \hat{z}_a + H_{p_i}\big(Z(s)\big)\tau_{x_i}(\vec{x})(-i\hbar\pa_s)+\dac{1}{2}H_{z_a z_b}\big(Z(s)\big)\Delta \hat{z}_a \Delta \hat{z}_b+\\
+H_{z_a p_i}\big(Z(s)\big)\tau_{x_i}(\vec{x})\Delta \hat{z}_a (-i\hbar\pa_s)+\dac{1}{2}H_{p_i p_j}\big(Z(s)\big)\Big(-\tau_{x_i x_j}(\vec{x})\hbar^2\pa_s+\tau_{x_i}\tau_{x_j}(-i\hbar\pa_s)^2\Big)+\\
+ \dac{1}{6} H_{z_a z_b z_c}\big(Z(s)\big) \Delta \hat{z}_a \Delta \hat{z}_b \Delta \hat{z}_c +\dac{1}{2}H_{z_a z_b p_k}\big(Z(s)\big) \tau_{x_k}(\vec{x})\Delta \hat{z}_a \Delta \hat{z}_b (-i\hbar \pa_s) +\\
+\dac{1}{2}H_{z_a p_j p_k}\big(Z(s)\big) \tau_{x_j}(\vec{x}) \tau_{x_k}(\vec{x})\Delta \hat{z}_a (-i\hbar \pa_s)^2+\dac{1}{6}H_{p_i p_j p_k}\big(Z(s)\big) \tau_{x_i}(\vec{x}) \tau_{x_j}(\vec{x}) \tau_{x_k}(\vec{x}) (-i\hbar \pa_s)^3+\\
+\dac{1}{2}H_{z_a p_j p_k}\big(Z(s)\big) (-i\hbar)\tau_{x_j x_k}(\vec{x}) \Delta \hat{z}_a (-i\hbar \pa_s)+\\
+\dac{1}{2}H_{p_i p_j p_k}\big(Z(s)\big) (-i\hbar)\tau_{x_i}(\vec{x})\tau_{x_j x_k}(\vec{x}) (-i\hbar \pa_s)^2+\dac{1}{6}H_{p_i p_j p_k}\big(Z(s)\big) (-i\hbar)^2\tau_{x_i x_j x_k}(\vec{x}) (-i\hbar \pa_s)+\hat{\Or}(\hbar^2).
\end{gathered}
\label{razras1}
\end{equation}

When we turn to the expanded space, the pseudo-differential operator of the form $\hat{A}(t)=A(\hat{z},\tau(\vec{x},t),t)$ is put into correspondence with the operator $\hat{A}(s,t)=A(\hat{z},s,t)$. Such correspondence is not unique but the result of our calculations does not depend on its particular choice.

The equation for the moments in the expanded space read:
\begin{equation}
\begin{gathered}
\dac{\pa}{\pa t}\llan A(\hat{z},s,t) \rran_{\Phi} = \llan \pa_t A(\hat{z},s,t)\rran_{\Phi}-i\hbar\llan \tau_{t}(\vec{x},t)\pa_s A(\hat{z},s,t)\rran_{\Phi}+\dac{i}{\hbar}\llan [H(\hat{z},t)[\Phi], A(\hat{z},s,t)] \rran_{\Phi}-\\
-\Lambda \llan [\breve{H}(\hat{z},t)[\Phi], A(\hat{z},s,t)]_+\rran_{\Phi}.
\end{gathered}
\label{razras2}
\end{equation}
Note that operators operators $H(\hat{z},t)[\Phi]$ and $\breve{H}(\hat{z},t)[\Phi]$ commute with $\pa_s$.

In view of the expansion \eqref{razras1}, the equations for up to the second order moments in the expanded space accurate to $\Or(\hbar^{3/2})$ are given by:
\begin{equation}
\begin{gathered}
{\color{blue} \dot{\mu}(s)}=-\Lambda\Big(2\breve{H}\mu(s)+2\breve{H}_{z_a}\Delta_a(s)+\breve{H}_{z_a z_b}\Delta_{ab}(s)+2\breve{H}_{p_j}\tau_{x_j}^{(0)}\Pi(s)\Big),\\
{\color{blue}\dot{\Delta}_{d}(s)}=J_{da} H_{z_a} \mu(s)-H_{p_k}\tau_{x_k}Z_{s,d}\mu(s)  -\tilde{\delta}_{dj}H_{p_k}\tau_{x_j x_k}^{(0)} \Pi(s) +J_{da}H_{z_a z_b}\Delta_{b}(s)+\\
+J_{da}H_{z_a p_k} \tau_{x_k}^{(0)} \Pi(s)- H_{z_a p_k} Z_{s,d}\llan\tau_{x_k}\rran - \tilde{\delta}_{dj}H_{z_a p_k}\tau_{x_k x_j}^{(0)} \Pi(s) -\\
-H_{p_k p_j}\tau_{x_k}^{(0)}\tau_{x_j}^{(0)}Z_{s,d}\Pi(s)+\dac{1}{2}J_{da}H_{z_a z_b z_c}\Delta_{bc}(s)+\dac{1}{2}H_{z_a z_b p_k} \tau_{x_k}^{(0)} Z_{s,d}\Delta_{ab}(s)-\\
 -\Lambda\left(2\breve{H} \Delta_{d}(s) + 2\breve{H}_{z_a} \Delta_{ad}(s) \right)-\dot{Z}_{d} \mu(s),\\
{\color{blue}\dot{\Delta}_{cd}(s)}=2J_{ca}H_{z_a}\Delta_{d}(s)+2H_{p_k}Z_{s,c}\llan \tau_{x_k} \Delta \hat{z}_d\rran+2J_{ca}H_{z_a z_b}\Delta_{bd}(s)-\\
-2H_{z_a p_k}\tau_{x_k}^{(0)}Z_{s,c}\Delta_{ad}(s)-\Lambda\left(2\breve{H}\Delta_{cd}(s)\right)-2\Delta_{c}(s) \dot{Z}_{d}(s),\\
{\color{blue}\dot{\Pi}(s)}=2\Lambda \breve{H} \Pi(s).
\end{gathered}
\label{cherna2}
\end{equation}
Here, we omitted the dependence of all functions on $t$ for brevity and we denoted $\tau_{x_{i_1}...x_{i_m}}^{(0)}(\vec{x},t)=\lim_{\hbar\to 0}\tau_{x_{i_1}...x_{i_m}}(\vec{x},t)$ in the operator sense, i.e. $\tau_{x_{i_1}...x_{i_m}}^{(0)}(\vec{x},t)=\tau_{x_{i_1}...x_{i_m}}(X(s,t),t)$. In the equations \eqref{cherna2}, we put $a,b,c,d=1,...,2n$ and $i,j,k=1,...,n$. Also, we have taken into consideration that $[\Delta \hat{z}_a,\Delta \hat{z}_b]=i\hbar J_{ab}$. The terms in the right-hand side of the equation for $\Delta_{cd}$ are assume to be symmetrizaed with respect to $c,d$ (the symmetrization is not written explicitly). The equations \eqref{cherna2} are written under assumption that the condition \eqref{expeq2} is satisfied.

The explicit form of coefficients in the equations \eqref{cherna2} within the sufficient accuracy for the given estimates reads as follows:
\begin{equation}
\begin{gathered}
\tau_{x_i}^{(0)}=\tau_{x_i}^{(0)}(s,t)=\dac{X_{s,i}(s,t)}{\vec{X}_s^2(s,t)},\\
\llan\tau_{x_i}\rran=\tau_{x_i}^{(0)}(s,t)\Bigg(\mu(s,t)+\dac{X_{ss,j}(s,t)\Delta_{j+n}(s,t)}{\vec{X}_s^2(s,t)}+\dac{X_{ss,j}(s,t)X_{ss,k}(s,t)\Delta_{j+n,k+n}(s,t)}{\left(\vec{X}^2_s(s,t)\right)^2}\Bigg)+\Or(\hbar^{3/2}),\\
\llan \tau_{x_i} \Delta \hat{z}_a\rran=\tau_{x_i}^{(0)}(s,t)\Bigg(\Delta_a(s,t)+\dac{X_{ss,j}(s,t)\Delta_{j+n,a}(s,t)}{\vec{X}_s^2(s,t)}\Bigg)+\Or(\hbar^{3/2}),\\
\end{gathered}
\label{podstan1}
\end{equation}

\begin{equation}
\begin{gathered}
H_{z_{a_1}...z_{a_m}}=H_{z_{a_1}...z_{a_m}}(s,t)=V_{z_{a_1}...z_{a_m}}\left(Z(s,t)\right)+\dil_{{\mathbb{D}}}dr \Bigg( W_{z_{a_1}...z_{a_m}}\left(Z(s,t),Z(r,t),t\right)\mu(r,t)+\\
+W_{z_{a_1}...z_{a_m} w_b}\left(Z(s,t),Z(r,t),t\right)\Delta_b(r,t)+W_{z_{a_1}...z_{a_m} w_b w_c}\left(Z(s,t),Z(r,t),t\right)\Delta_{bc}(r,t)\Bigg)+\Or(\hbar^{3/2}),
\end{gathered}
\label{podstan2}
\end{equation}
and the similar to \eqref{podstan2} expansion for $\breve{H}_{z_{a_1}...z_{a_m}}$.

\bibliography{lit1}

\end{document}